# CONVEX BASES OF PBW TYPE FOR QUANTUM AFFINE ALGEBRAS

JONATHAN BECK

**Abstract.** This note has two purposes. First we establish that the map defined in [L, §40.2.5 (a)] is an isomorphism for certain admissible sequences. Second we show the map gives rise to a convex basis of Poincaré–Birkhoff–Witt (PBW) type for $\mathbf{U}^+$, an affine untwisted quantized enveloping algebra of Drinfel'd and Jimbo. The computations in this paper are made possible by extending the usual braid group action by certain outer automorphisms of the algebra.

**Introduction.** One of the basic difficulties in working with the quantized enveloping algebras is that they are deformations of a given universal enveloping algebra rather than the underlying Kac–Moody Lie algebra. Since a linear basis is no longer obtained using the Poincaré–Birkhoff–Witt theorem, a first task is to construct a basis of the algebra $\mathbf{U}^+$. A PBW type basis of $\mathbf{U}^+$ is formed by ordered monomials in root vectors $E_\alpha$, where each $E_\alpha$ specializes at 1 (in the sense of [L3]) to an $\alpha$–root vector of $\widehat{\mathfrak{g}}$.

This paper treats the problem of finding a PBW type basis when the Cartan datum is the affine extension of a finite Cartan datum. In the case when the underlying type is $\mathfrak{sl}_2$, the basis given here is identical to that of [Da], [LSS]. This basis completes the construction proposed in [L §40.2]. The principal missing part of that construction is an explicit description of the imaginary root space, and that is described here. We define a convex basis which is formed by monomials in certain root vectors of $\mathbf{U}^+$ multiplied in a predetermined total order on the root system.

The convexity property, which appeared in the work of [L–S] for the finite type case, means that the $q$–commutator of two root vectors, $E_\alpha$ and $E_\beta$, consists of monomials formed only from root vectors between $\alpha$ and $\beta$ in the order. This basis should be useful for a variety of applications. For example, one can explicitly construct the universal R–matrix in terms of the braid group action by a direct extension of the work of [LSS]. This construction uses braid group operators arising from the lattice of translations in the extended affine Weyl group. In the works ([K–T], [K–T2]) convex bases and the universal R–matrix are also contructed, although the braid group is not used and proofs are not given.

**Notation.** The notation follows that in [L]. Let $\mathbf{U}$ be the quantized enveloping algebra corresponding to an untwisted affine Cartan datum $(\tilde{I}, \cdot)$. Denote its Weyl group by $\tilde{W}$, a Coxeter group on a set of simple reflections $S = \{s_0, s_1, \ldots, s_n\}$.

Let $Q$ be the normal subgroup of $\tilde{W}$ consisting of all elements with finitely many conjugates. Let $\Omega$ be the group of automorphisms of $(\tilde{W}, \tilde{I})$ whose restriction to $Q$ is conjugation by some element of $\tilde{W}$. $\Omega$ is a finite group in correspondence with a certain subgroup of automorphisms of the graph of $(\tilde{I}, \cdot)$ (see [B]). The extended affine Weyl group is defined as $W = \Omega \ltimes \tilde{W}$, where the product is given by $(\tau, w)(\tau', w') = (\tau\tau', \tau'^{-1}(w)w')$. The length function of $\tilde{W}$ extends to $W$ by setting $l(\tau w) = l(w)$ for $\tau \in \Omega$. Fix an index $i_0 \in \tilde{I}$ so that the simply connected root datum $(\tilde{Y}, \tilde{X}, \langle,\rangle, \ldots)$ of $\tilde{I}$ restricts to a root datum $(Y, X, \langle,\rangle, \ldots)$ of $(\tilde{I} \setminus \{i_0\}, \cdot)$, the underlying finite type Cartan datum of $(\tilde{I}, \cdot)$.

Let $W_0$ be the Weyl group of $I = \tilde{I} \setminus \{i_0\}$. Then $W \cong X \rtimes W_0$ and $X$ characterized as being the subgroup of elements of $W$ with finitely many conjugates. It is known that $X \supset Q$ and $X/Q \cong \Omega$. Let $\{\omega_i\}_{i \in I} \subset X = \text{Hom}(Y, \mathbb{Z})$ be the dual basis of $Y$. Let $P^{++}$ be the semigroup in $X$ generated by $\omega_i$. Then $P^{++}$ has the properties:

$$P^{++} = \{x \in X \mid l(s_i x) = l(x) + 1,\ 1 \leq i \leq n\}$$

(*) $\quad l(xy) = l(x) + l(y),\ \text{for } x, y \in P^{++}.$

The orbit $\tilde{\mathcal{R}}$ of $\tilde{I}$ under $\tilde{W}$ consists of the real coroots. Denoting by $\mathcal{R} \subset Y$ the coroot set of $(Y, X, \langle,\rangle, \ldots)$ there is a well-known correspondence between the following sets.

$$\tilde{\mathcal{R}}^+ \leftrightarrow \{(\check{\alpha}, k) \mid \check{\alpha} \in \mathcal{R}, k > 0\} \cup \{(\check{\alpha}, 0) \mid \check{\alpha} \in \mathcal{R}^+\},$$
$$\tilde{\mathcal{R}}^- \leftrightarrow \{(\check{\alpha}, k) \mid \check{\alpha} \in \mathcal{R}, k < 0\} \cup \{(\check{\alpha}, 0) \mid \check{\alpha} \in \mathcal{R}^-\},$$

such that $\tilde{\mathcal{R}} = \tilde{\mathcal{R}}^+ \cup \tilde{\mathcal{R}}^-$.

We define the braid group of $W$ on generators $T_w$, $w \in W$ with relations $T_w T_{w'} = T_{ww'}$ when $l(ww') = l(w) + l(w')$. Write $\tau$ for $T_\tau$. We extend the symmetries on $\mathbf{U}$ to correspond to the braid group of $W$. For $\tau \in \Omega$ this is done by defining $\tau E_i = E_{\tau(i)}$, $\tau F_i = F_{\tau(i)}$, and $\tau K_i = K_{\tau(i)}$, $i \in \tilde{I}$.

Let $w \in W$. Given a reduced presentation $w = s_{i_1} s_{i_2} \ldots s_{i_N}$ define the *initial set* of $w$ to be:

$$I_w = \{\beta_k \mid \beta_k = s_{i_1} s_{i_2} \ldots s_{i_{k-1}}(\alpha_{i_k}),\ 1 \leq k \leq N\},$$

and the *terminal set* to be:

$$E_w = I_{w^{-1}} = \{\beta_k \mid \beta_k = s_{i_N} s_{i_{N-1}} \ldots s_{i_{k+1}}(\alpha_{i_k}),\ 1 \leq k \leq N\}.$$

$I_w$ is independent of the choice of reduced expression of $w$ and is characterized as the set of $\check{\alpha} \in \tilde{\mathcal{R}}^+$ such that $w^{-1}(\check{\alpha}) \in \tilde{\mathcal{R}}^-$.

§1 **Convex PBW bases.**

Let $x \in Q$ such that $\langle i, x \rangle > 0$ for $i \in I$. Fix a reduced presentation of $x = s_{i_1} s_{i_2} \ldots s_{i_N}$. By property (*) the following sequence is admissible. For $k \in \mathbb{Z}$ let $i_k = i_{k \bmod(N)}$,

(0) $\quad\quad\quad\quad\quad\quad\quad\quad \mathbf{h} = (\ldots i_{-1}, i_0, i_1, i_2 \ldots).$

**Lemma 1.** *Let $r > 0$.*
  (a) $I_{x^r} = \{(-\check{\alpha}, k) \mid \check{\alpha} \in \mathcal{R}^+,\ 1 \leq k \leq r\langle \check{\alpha}, x \rangle\}$,
  (b) $E_{x^r} = \{(\check{\alpha}, k) \mid \check{\alpha} \in \mathcal{R}^+,\ 0 \leq k \leq r\langle \check{\alpha}, x \rangle - 1\}$.

*Proof.* The terminal set of an element $w \in W$ is the set of positive coroots $w$ maps to negative coroots. $x$ acts on the set of positive real coroots by $x(\check{\alpha}, k) = (\check{\alpha}, k - \langle \check{\alpha}, x \rangle)$. This establishes (b). (a) is similar. $\square$

Let $\mathbf{P}$ be the set of elements $y \in \mathbf{U}^+$ for which $T_{i_s}^{-1} T_{i_{s-1}}^{-1} \ldots T_{i_1}^{-1} y \in \mathbf{U}^+$, $T_{i_r} T_{i_{r+1}} \ldots T_{i_0} y \in \mathbf{U}^+$ for $s > 0, r < 0$.

Let $y \in \mathbf{P}$. For any sequence $\mathbf{c} = (\ldots c_{-2}, c_{-1}, c_0, c_1, \ldots)$, $c_i \in \mathbb{N}$, where almost all $c_i = 0$ define

$$L(\mathbf{h}, \mathbf{c}, y) = (E_{i_0}^{(c_0)} T_{i_0}^{-1}(E_{i_{-1}}^{(c_{-1})}) T_{i_0}^{-1} T_{i_{-1}}^{-1}(E_{i_{-2}}^{(c_{-2})}) \ldots) \times y \times (\ldots T_{i_1}(E_{i_2}^{(c_2)}) E_{i_1}^{(c_1)}).$$

Let $\mathbf{U}^+(>)$ (resp. $\mathbf{U}^+(<)$) be the subspace of $\mathbf{U}^+$ spanned by the elements $E_{i_0}^{(c_0)} T_{i_0}^{-1}(E_{i_{-1}}^{(c_{-1})}) T_{i_0}^{-1} T_{i_{-1}}^{-1}(E_{i_{-2}}^{(c_{-2})}) \ldots$ (resp. $\ldots T_{i_1}(E_{i_2}^{(c_2)}) E_{i_1}^{(c_1)}$) for various $\mathbf{c}$. Notice that by [L 40.2.1] $\mathbf{U}^+(>)$ and $\mathbf{U}^+(<)$ are independent of the reduced expression for $x$ chosen.

By [L 40.2.5 (a)] we have:

(1) $$\mathbf{U}^+(>) \otimes \mathbf{P} \otimes \mathbf{U}^+(<) \to \mathbf{U}^+$$

given by multiplication is an injective map. We describe $\mathbf{P}$ for the admissible sequence $(0)$.

Define the imaginary root vectors $E_{k\delta}^i$, $1 \leq i \leq n$, $k \in \mathbb{N}$ by:

$$E_{k\delta}^i = q_i^{-2} E_i T_{\omega_i}^k(K_i^{-1} F_i) - T_{\omega_i}^k(K_i^{-1} F_i) E_i.$$

**Lemma 2.** *Let $1 \leq i \leq n$, $k > 0$. Then $E_{k\delta}^i \in \mathbf{P}$.*

*Proof.* We demonstrate this for a particular reduced expression of $x$, from which the lemma will follow independently of the reduced presentation. Write $x = \omega_1^{l_1} \omega_2^{l_2} \ldots \omega_n^{l_n}$ and fix a reduced presentation of $x$ which is a concatenation of reduced presentations of the $\omega_i$ in the given order. Note that $\omega_i \in \Omega \ltimes \tilde{W}$, $1 \leq i \leq n$, but since $x \in Q$ we can collect all elements $\tau \in \Omega$ on the left and they will cancel, leaving an element of $Q$ which has a reduced expression in terms of simple reflections. Since for $\tau \in \Omega$, $\tau(u) \in \mathbf{U}^+ \leftrightarrow u \in \mathbf{U}^+$, we can work with reduced presentations of $\omega_i$.

Since $T_x(E_{k\delta}^i) = E_{k\delta}^i$ (see [Be], [Da]) ($1 \leq i \leq n$) it is sufficient to check that:

(2) $$T_{j_r} T_{j_{r+1}} \ldots T_{j_d}(E_{k\delta}^i) \in \mathbf{U}^+,$$
$$T_{j_{r-1}}^{-1} \ldots T_{j_1}^{-1} \tau^{-1}(E_{k\delta}^i) \in \mathbf{U}^+, \quad 1 \leq r \leq d.$$

where $\tau s_{j_1} \ldots s_{j_d}$ is a reduced presentation of some $\omega_j$. Further, since $T_{\omega_j}(E_{k\delta}^i) = E_{k\delta}^i$ the second expression equals the first and it is sufficient to check the first.

If $j = i$ then necessarily $j_d = i$ and

$$T_i(E^i_{k\delta}) = T_i(q_i^{-2} E_i T^k_{\omega_i}(K_i^{-1} F_i) - T^k_{\omega_i}(K_i^{-1} F_i) E_i)$$
$$= q_i^{-2} T_{\omega'_i}(E_i) T_i T^{k-1}_{\omega_i}(K_i^{-1} F_i) - T_i T^{k-1}_{\omega_i}(K_i^{-1} F_i) T_{\omega'_i}(E_i).$$

The calculation of the last equality is found in [Be]. The lemma now follows by [L 40.1.2] and the consideration that $l(\omega_i s_i \omega_i) = 2l(\omega_i) - 1$ [L2, Lemma 2.3]. If $j \neq i$ the lemma is clear using [L 40.1.2] since $l(\omega_j s_i) = l(\omega_j) + 1$ and $l(\omega_j \omega_i) = l(\omega_j) + l(\omega_i)$.

It remains to show the lemma for any reduced presentation of $x$. Such a presentation can be transformed to the above one by braid relations alone. Since the braid relations preserve the length of a reduced expression, the result follows from [L 40.1.2].

It is convenient to renormalize the imaginary root vectors by the functional equation:

$$1 + (q_i - q_i^{-1}) \sum_{k \geq 0} E^i_{k\delta} u^k = \exp\left((q_i - q_i^{-1}) \sum_{k=1}^{\infty} \tilde{E}^i_{k\delta} u^k\right).$$

Index the $\tilde{E}^i_{k\delta}$ by $S = \{1, 2, \ldots, n\} \times \mathbb{N}$ and for $s = (i, k) \in S$ write $\tilde{E}_s$ for $\tilde{E}^i_{k\delta}$. Fix an order on $S$ and consider the subset of $\mathbf{P}$,

$$\mathbf{X} = \left\{ \prod_{s \in S} \tilde{E}^{c_s}_s \mid c_s \in \mathbb{N}, c_s = 0 \text{ for almost all } s \right\}$$

where the product is taken in a fixed order. Then $\mathbf{X} \subset \mathbf{P}$. By [Be, Prop. 6.1] we have:

**Proposition 3.** *Let $y, y' \in \mathbf{X}$, $\mathbf{c} = (c_i), \mathbf{c}' = (c'_i)$, almost all $c_i, c'_i = 0$. Let $t = \prod_{i \in \tilde{I}} K_i^{m_i}$, $m_i \in \mathbb{Z}$.*

  (a) *The expressions $L(\mathbf{h}, \mathbf{c}, y)$ form a linear basis of the $\mathbb{Q}(q)$–vector space $\mathbf{U}^+$.*
  (b) *The expressions $L(\mathbf{h}, \mathbf{c}, y) \times t \times \Omega(L(\mathbf{h}, \mathbf{c}', y'))$ form a linear basis of the $\mathbb{Q}(q)$–vector space $\mathbf{U}$.*

where $\Omega$ is the standard anti–involution of $\mathbf{U}$.

Further, since $\mathbf{U}^+(>) \otimes \mathbf{P} \otimes \mathbf{U}^+(<)$ imbeds into $\mathbf{U}^+$ we conclude:

**Corollary 4.**
  (a) $\mathbf{X}$ *is a basis of the subalgebra $\mathbf{P}$ of $\mathbf{U}^+$.*
  (b) $\mathbf{U}^+ \cong \mathbf{U}^+(>) \otimes \mathbf{P} \otimes \mathbf{U}^+(<)$.
  (c) $\mathbf{U} \cong \mathbf{U}^+(>) \otimes \mathbf{P} \otimes \mathbf{U}^+(<) \otimes \mathbf{U}^0 \otimes \Omega(\mathbf{U}^+(<)) \otimes \Omega(\mathbf{P}) \otimes \Omega(\mathbf{U}^+(>))$.

We recall some facts about the quantum affine algebras (see [Be]). Let $x^+_{ik} = T^{-k}_{\omega_i}(E_i)$, for $k \geq 0$, $x^-_{ik} = T^k_{\omega_i}(-K_i^{-1} F_i)$ for $k > 0$. Note that $x^+_{ik} \in \mathbf{U}^+$ for $k \geq 0$, $x^-_{ik} \in \mathbf{U}^+$ for $k > 0$.

The following commutation relations hold in $\mathbf{U}^+$.

(3)
$$[\tilde{E}^i_{k\delta}, \tilde{E}^j_{l\delta}] = 0, \quad 1 \leq i, j \leq n, \ k, l > 0,$$

$$[\tilde{E}^i_{k\delta}, x^+_{jl}] = \frac{(\operatorname{sgn}(a_{ij}))^k [k a_{ij}]_i}{k} x^+_{j,l+k}, \quad l \geq 0$$

$$[\tilde{E}^i_{k\delta}, x^-_{jl}] = \frac{(\operatorname{sgn}(a_{ij}))^k [k a_{ij}]_i}{k} x^-_{j,l+k}, \quad l > 0$$

Define $x_{i,-k}^- = \Omega(x_{ik}^+)$ for $k \geq 0$, $x_{i,-k}^+ = \Omega(x_{ik}^-)$, $k > 0$. Let $\tilde{F}_{k\delta}^i = \Omega(\tilde{E}_{k\delta}^i)$ for $k > 0$. We now consider the following subalgebras of $\mathbf{U}$.

$$A_> = \{u \in \mathbf{U} \mid (T_x)^k u \in \mathbf{U}^-\mathbf{U}^0, k \gg 0\}$$
$$A_< = \{u \in \mathbf{U} \mid (T_x)^k u \in \mathbf{U}^-\mathbf{U}^0, k \ll 0\}$$

Note that $\mathbf{U}^+(<) \subset A_<$, $\mathbf{U}^+(>) \subset A_>$.

**Lemma 5.**
  (a) $\mathbf{U}^+(>) = A_> \cap \mathbf{U}^+$.
  (b) $\mathbf{U}^+(<) = A_< \cap \mathbf{U}^+$.

*Proof.* $\mathbf{U}^+(>) \subset A_> \cap \mathbf{U}^+$ is clear. Now use Prop. 3. Let $u \in (A_> \cap \mathbf{U}^+) \setminus \mathbf{U}^+(>)$. By Prop. 3, $u = \sum c_{g_1,p,g_2} g_1 \cdot p \cdot g_2$ where $g_1 \in \mathbf{U}^+(>)$, $p \in \mathbf{P}$, $g_2 \in \mathbf{U}^+(<)$. By assumption some $c_{g_1,p,g_2} \neq 0$ for $p$ or $g_2$ not equal 1. Fix $k > 0$ so that $(T_x)^k(u) \in \mathbf{U}^-\mathbf{U}^0$. By definition

$$(T_x)^k(\sum c_{g_1,p,g_2} g_1 \cdot p \cdot g_2) = \sum c_{g_1,p,g_2}(T_x)^k(g_1) \cdot (T_x)^k(p \cdot g_2).$$

The last expression is a sum in PBW monomials for $\mathbf{U}$ (coming from Corollary 4 (c)). However, $(T_x)^k(p \cdot g_2) \in \mathbf{U}^+$. It follows $A_> \cap \mathbf{U}^+ = \mathbf{U}^+(>)$. (b) is similar.

Note that Lemma 5 implies that $\mathbf{U}^+(>)$, $\mathbf{U}^+(<)$ are subalgebras of $\mathbf{U}$.

**Lemma 6.**
  (a) $[\mathbf{P}, \mathbf{U}^+(>)] \subset \mathbf{U}^+(>)$,
  (b) $[\mathbf{P}, \mathbf{U}^+(<)] \subset \mathbf{U}^+(<)$.

*Proof.* We prove (a). By the previous Lemma it is sufficient to demonstrate that $[\mathbf{P}, A_>] \subset A_>$. Let $N^+$ (resp. $N^-$) be the subalgebra of $\mathbf{U}$ generated over $\mathbb{C}(q)$ by $x_{ik}^+$, $k \in \mathbb{Z}$ (resp. $x_{ik}^-$, $k \in \mathbb{Z}$). Let $H^+$ (resp. $H^-$) be the subalgebra generated by the $\tilde{E}_{k\delta}^i$ (resp. $\tilde{F}_{k\delta}^i$). We show that $A_>$ is generated as a subalgebra over $\mathbb{C}(q)$ by $N^+$, $H^-$ and $\mathbf{U}^0$. Certainly these are subalgebras of $A_>$. It is known that $U_q = N^+ \otimes H^- \otimes \mathbf{U}^0 \otimes H^+ \otimes N^-$. Let $y \in A_>$. Write $y = \sum_{s \in S} a_s n_s^+ \times h_s^- \times t \times h_s^+ \times n_s^-$, where $n_s^\pm, h_s^\pm, t$ are elements of given bases of $N^\pm$, $H^\pm$ and $\mathbf{U}^0$ respectively. Here each $a_s \in \mathbb{Q}(q)$ and $S$ is some finite index set for the summation. Fix $k'$ so that for $k > k'$, $T_x^k(y) \in \mathbf{U}^-\mathbf{U}^0$. Now by the definitions of the $x_{ik}^\pm$ it is possible to fix $k''$ large enough so that for $k > k''$ we have $T_x^k(n_s^+) \in \mathbf{U}^-\mathbf{U}^0$, $T_x^k(n_s^-) \in \mathbf{U}^+\mathbf{U}^0$. Note that $T_x(h) = h$ for all $h \in H$. By considering $k > k', k''$ and using triangular decomposition it follows that $n_s^- = 1$, $h_s^+ = 1$ for all $s \in S$.

Consider the basis of $\mathbf{U}^+$ consisting of the elements

$$L(\mathbf{h}, \mathbf{c}, y), \ y \in \mathbf{X}$$

$\mathbf{h}, \mathbf{c}$ as above. Let $\alpha_i = i' \in X$, $1 \leq i \leq n$. For $k \leq 0$ let $\beta_k = s_{i_0} \ldots s_{i_{k-1}}(\alpha_{i_k})$, and for $k > 0$ let $\beta_k = s_{i_1} \ldots s_{i_{k-1}}(\alpha_{i_k})$. Let $\delta$ be the image in $X$ of the unique element of $\mathbb{N}[I]$ with relatively prime coordinates such that $|\delta| \cdot |i| = 0$, for $i \in \tilde{I}$. In other words, let $\delta = \theta + \alpha_{i_0}$ where $\theta$ is the highest root of $W_0 \cdot \{\alpha_i\}_{i \in I}$. Consider the total order on the affine root system:

$$\beta_0 < \beta_{-1} < \beta_{-2} < \cdots < 2\delta < \delta \cdots < \beta_3 < \beta_2 < \beta_1 \tag{4}$$

We introduce real root vectors by defining

$$E_{\beta_k} = T_{i_0}^{-1} \ldots T_{i_{k+1}}^{-1}(E_{i_k}), \quad k \leq 0$$
$$E_{\beta_k} = T_{i_1} T_{i_2} \ldots T_{i_{k-1}}(E_{i_k}), \quad k > 0$$

For the imaginary root $k\delta$, order the imaginary root vectors $\tilde{E}_{k\delta}^i$, $(1 \leq i \leq n)$ arbitrarily. Then together with (4) we have introduced a total ordering on a set of root vectors of $\mathbf{U}^+$.

**Proposition 7.** Let $E_\beta > E_\alpha$.

$$E_\beta E_\alpha - q^{|\alpha| \cdot |\beta|} E_\alpha E_\beta = \sum_{\alpha < \gamma_1 < \cdots < \gamma_n < \beta} c_{\vec{\gamma}} E_{\gamma_1}^{a_1} \ldots E_{\gamma_n}^{a_n}.$$

where $c_{\vec{\gamma}} \in \mathbb{C}(q)$ for $\vec{\gamma} = (\gamma_1, \gamma_2, \ldots, \gamma_n)$.

*Proof.* The proof is a case by case analysis as in [L–S]. Consider the case where $E_\beta$ and $E_\alpha$ are real root vectors. Using the PBW basis, write

$$E_{\beta_k} E_{\beta_{k'}} = \sum c(q)_{\vec{\gamma}} E_{\gamma_1}^{a_1} E_{\gamma_2}^{a_2} \ldots E_{\gamma_n}^{a_n} \tag{5}$$

where the order on $\gamma_1, \gamma_2, \ldots, \gamma_n$ is as in (4).

Case (1): $k' < k < 0$. Assume $\gamma_1 = \beta_{k''}$ where $k'' < k'$. Apply $T_{i_{k''}} T_{i_{k''+1}} \ldots T_{i_0}$ to both sides of (5). One obtains an expression of the form

$$T_{i_{k''-1}}^{-1} \ldots T_{i_{k'+1}}^{-1}(E_{i_{k'}}) T_{i_{k''-1}}^{-1} \ldots T_{i_{k+1}}^{-1}(E_{i_k}) \in \sum_{a_1} c(q)_{\gamma_1^{a_1} \vec{\gamma}} F_{i_{k''}}^{a_1} K_{i_{k''}}^{a_1}(\mathbf{U}^+) + \mathbf{U}^+.$$

This implies (using triangular decomposition) that for each $a_1$, $c(q)_{\gamma_1^{a_1} \vec{\gamma}} = 0$, which contradicts the assumption that $k'' < k'$. One argues similarly if $\gamma_1 = \beta_{k''}$ for $k'' > k$. Therefore,

$$E_{\beta_k} E_{\beta_{k'}} - a E_{\beta_{k'}} E_{\beta_k} = \sum_{\alpha < \gamma_1 < \cdots < \gamma_r < \beta} c(q)_{\vec{\gamma}} E_{\gamma_1}^{a_1} E_{\gamma_2}^{a_2} \ldots E_{\gamma_r}^{a_r}. \tag{6}$$

Applying $T_{i_k} T_{i_{k-1}} \ldots T_{i_0}$ to both sides of (5) we obtain

$$- F_{i_k} K_{i_k} T_{i_{k-1}}^{-1} \ldots T_{i_{k'+1}}^{-1}(E_{i_{k'}}) + a T_{i_{k-1}}^{-1} \ldots T_{i_{k'-1}}^{-1}(E_{i_{k'}}) F_{i_k} K_{i_k}$$
$$= (-q^{|\alpha| \cdot |\beta|} [F_{i_k}, T_{i_{k+1}}^{-1} \ldots T_{i_{k'+1}}^{-1}(E_{i_{k'}})] + (a - q^{|\alpha| \cdot |\beta|}) T_{i_{k+1}}^{-1} \ldots T_{i_{k'+1}}^{-1}(E_{i_{k'}}) F_{i_k}) K_{i_k}$$
$$= \sum c(q)_{\vec{\gamma}} E_{\gamma_1}^{a_1} E_{\gamma_2}^{a_2} \ldots E_{\gamma_r}^{a_r} \in \mathbf{U}^+.$$

Since the left hand side is also in $\mathbf{U}^+$ it follows that $a = q^{|\alpha|\cdot|\beta|}$.

Case (2): $k' < 0 < k$. This is similar.

Case (3): Assume $\beta = r\delta$, $\alpha = \beta_{k'}$, $k' \leq 0$. By Lemma 6 we have

$$E_\beta E_{\beta_{k'}} - E_{\beta_{k'}} E_\beta = \sum c(q)_\beta E_{\beta_{k_1}}^{a_1} E_{\beta_{k_2}}^{a_2} \ldots E_{\beta_{k_n}}^{a_n}$$

where for $1 \leq i \leq n$, $k_i \leq 0$. The convexity is checked by verifying $k < k'_i$ for $1 \leq i \leq n$. This follows from triangular decomposition as before.

If $\beta = r\delta$, $\alpha = \beta_{k'}, k' > 0$, the situation is similar to the previous case.

*Remark.* For $U_q(\widehat{\mathfrak{sl}_2})$ there are two admissible sequences, either $i_k = k(\mathrm{mod}\ 2)$ or $i_k = k+1(\mathrm{mod}\ 2)$. Both of these are of the form above and are obtained by considering the affine Cartan datum $(\{0,1\},\cdot)$ together with an underlying finite Cartan datum of the same type. In the first case one obtains the above description of $\mathbf{P}$ when $I = \tilde{I} \setminus \{0\}$ and in the second case when $I = \tilde{I} \setminus \{1\}$. In cases other than $\widehat{\mathfrak{sl}_2}$ not all admissible sequences are of the type considered here. For example, one can pick an arbitrary concatenation of the fundamental weights $\omega_i \in W$. In this case the results here hold without modification if each $\omega_i$ ($1 \leq i \leq n$) appears an infinite number of times to the left and right of $i_0$.

**Acknowledgement.** I am grateful to V. Kac and the MIT mathematics department for support during the course of this work.